\documentclass[amsmath,11pt]{article}

\usepackage{amssymb,amssymb,epsfig,bm}
\usepackage{caption}
\usepackage{subcaption}

\date{\empty}

\begin{document}

\title{\bf Cosmological peculiar velocities in general relativity?}
\author{Christos G. Tsagas\\ {\small ${}^1$Section of Astrophysics, Astronomy and Mechanics, Department of Physics}\\ {\small Aristotle University of Thessaloniki, Thessaloniki 54124, Greece}\\ {\small ${}^2$Clare Hall, University of Cambridge, Herschel Road, Cambridge CB3 9AL, UK}}

\maketitle

\vspace{-20pt}

\begin{abstract}
Cosmological peculiar velocities have traditionally been studied within the framework of Newtonian theory. Around the turn of the century, a few quasi-Newtonian analyses appeared in the literature, but led to equations and results identical to those of the purely Newtonian approach~\cite{M}. More recently, a series of studies introduced a relativistic treatment of the peculiar-velocity problem, criticising the quasi-Newtonian approach as effectively Newtonian in nature~\cite{T1}. These works also reported a linear growth-rate of $v \propto t$ for peculiar velocities, in contrast to the slower Newtonian/quasi-Newtonian scaling of $v \propto t^{1/3}$. In a manuscript uploaded to the archives a few days ago~\cite{CM}, the authors defended their earlier quasi-Newtonian work and criticised the more recent relativistic treatments. However, the limitations of the quasi-Newtonian approach are not a new concern, but they have been noted at least since~\cite{EMM}. There, it was clearly stated that the quasi-Newtonian approximation leads to Newtonian-like equations and results, and readers were cautioned against applying it to large-scale cosmological studies. Given that one of the authors of~\cite{CM} was also a coauthor of~\cite{EMM}, the self-contradiction is evident. The relativistic analyses have, in fact, confirmed the concerns of~\cite{EMM}, clarified the underlying issues and shown how they can be resolved. Motivated by~\cite{CM}, we present a critical comparison of the two approaches and in the process identify several internal inconsistencies in that manuscript.
\end{abstract}

\section{Introduction}\label{sI}\vspace{-5pt}
Peculiar velocities are ubiquitous in the universe, with numerous surveys confirming their presence. In fact, an increasing number of these studies report bulk peculiar flows whose scales and amplitudes exceed - sometimes significantly - the expectations of the standard model.

Predictions within the standard model are largely based on Newtonian gravity. Around the turn of the century, a few quasi-Newtonian treatments also appeared in the literature. Although these studies begin with a relativistic profile, they impose stringent constraints upon the perturbed spacetime, namely vanishing vorticity, zero shear, and the absence of gravitational waves. As a result, the 4-acceleration assumes a Newtonian-like form and its time-evolution is determined through a non-uniquely specified ansatz. Consequently, the analysis effectively reduces to a purely Newtonian one, leading to results identical to the Newtonian~\cite{M}. This is not an isolated incident. The relativistic treatment of the Zeldovich approximation also reproduced the exact Newtonian results, when the strict quasi-Newtonian constraints were imposed~\cite{ET}.

The limitations of the quasi-Newtonian approximation have been recognized at least since~\cite{EMM}. There, on pages 150–151, the authors clearly note that the stringent restrictions imposed by the approach on the perturbed spacetime lead to Newtonian-like equations and results. They also go further, cautioning readers against applying the quasi-Newtonian method to large-scale cosmological studies (see page 151 in~\cite{EMM}); a similar warning also appears on page 20. What was missing at the time, however, was a direct comparison between the quasi-Newtonian and a proper relativistic analysis of the same problem. The relativistic studies of linear peculiar velocities have since confirmed the concerns raised in~\cite{EMM}, identified the origin of the problem, revealed its extent and demonstrated how it can be resolved (e.g.~see~\cite{T1} and \S~\ref{ssRA} here).

In studies of peculiar velocities, the principal limitation of the quasi-Newtonian approximation does not lie in its constraints on the linear vorticity and shear. Rather, the key issue is that these restrictions force both the 4-acceleration and its time evolution to take on Newtonian forms. As a consequence, a genuinely general-relativistic effect, namely the gravitational contribution of the \textit{peculiar flux}, is inadvertently bypassed and not accounted for.

What is the peculiar flux? To understand this, recall that peculiar motions are matter in motion and moving matter implies nonzero energy flux. In relativity, unlike Newtonian gravity, energy fluxes contribute to the gravitational field. The resulting flux-effects persist even at the linear level and appear in all equations governing the evolution and the consequences of peculiar flows in cosmology. Clearly, a study of peculiar velocities that neglects such effects cannot be regarded as relativistic. Nevertheless, the quasi-Newtonian approach bypasses the gravitational input of the peculiar flux due to its strict constraints. The implications of this omission remained unrecognized, however, until the first fully relativistic treatment was introduced a few years ago~\cite{TT}.

Without accounting for the flux-effects, the quasi-Newtonian treatment employs only two differential equations to monitor the peculiar-velocity field, just as in the purely Newtonian studies. By contrast, without imposing any linear constraints and including the flux effects, the relativistic analysis arrived at a (considerably different) set of three equations. This led to a faster linear growth-rate for the peculiar-velocity field, that is to $v\propto t$~\cite{T1}, compared to the Newtonian/quasi-Newtonian $v\propto t^{1/3}$ rate. The mathematics are straightforward to follow and the underlying physical interpretation transparent to comprehend. \textit{In fact, had the energy-conservation law been employed in~\cite{M}, that study could also have arrived at the relativistic expression (\ref{RAa}) for the 4-acceleration, instead of the Newtonian-like (\ref{QNAa}a) (see \S~\ref{ssQNA} and \S~\ref{ssRA} below). Then, the present exchange would have been completely unnecessary.}

Relative motions have long been known to interfere with observations, potentially “contaminating” the data and leading to their misinterpretation. Historically, such effects have given rise to misconceptions, some of which were significant and persisted for centuries, if not millennia. A well-known example is the apparent motion of the planets, which once required increasingly complex systems of epicycles, but was ultimately resolved by adopting the correct frame of reference. Cosmology is not a priori immune to contaminations from relative motion and analogous effects may likewise lead to erroneous conclusions. \textit{For instance, without accounting for their position and kinematics, ``unsuspecting'' observers residing within locally contracting bulk flows, can be misled into inferring that the deceleration parameter of the universe has recently turned negative. Intuitively, these observers mistake the local contraction of their bulk flow for global acceleration of the surrounding universe.} In this case, both the apparent sign-change and the inferred cosmic acceleration may not be physical, but rather artefacts of relative motion~\cite{T2}, much like the perceived retrograde motion of the planets was.

Testing this possibility requires searching for the characteristic “trademark signature” of relative motion in the data, namely for an apparent (Doppler-like) dipolar anisotropy. Furthermore, the magnitude of this dipole should decrease with increasing redshift. In other words, to unsuspecting observers, the universe would appear to accelerate more rapidly in one direction on the sky and more slowly in the opposite (antipodal) direction. For more than a decade, dipolar patterns have been reported in the sky-distribution of the deceleration parameter. It was not until~\cite{CMRS}, however, that the dipole identified in the JLA catalogue was attributed to our peculiar motion relative to the rest-frame of the Cosmic Microwave Background (CMB). A few years later, a similar dipole was detected in the Pantheon+ dataset, notably with a magnitude that decreases with redshift~\cite{SRST}. Around the same time, a complementary analysis by~\cite{CRMC} reported a dipole in the distribution of $\Omega_{\Lambda}$, also consistent with the effects of our peculiar motion.

In~\cite{CM}, the authors criticised these results. However, in doing so, they unwittingly identified themselves with the unsuspecting observers who neglect the effects of their relative-motion. Despite inhabiting a universe permeated by peculiar flows and despite failing to account for them, such (unsuspecting) observers remain confident in the validity of their conclusions (see \S~\ref{sssRCs2} below).\vspace{-10pt}

\section{Relativistic vs quasi-Newtonian study}\label{sRQN}\vspace{-5pt}
In this section we will critically review the key differences between the quasi-Newtonian and the relativistic analyses of cosmological peculiar velocities. We will also review the criticisms of~\cite{CM}.\vspace{-10pt}

\subsection{The initial setup}\label{ssIS}\vspace{-5pt}
Studies involving peculiar velocities require tilted spacetimes that allow for two families of relatively moving coordinate systems located at every spacetime event. The first is the global rest-frame of the CMB and has 4-velocity $u_a$. The second frame, with 4-velocity $\tilde{u}_a$, is local and follows the peculiar motion of the matter. Assuming that the peculiar velocities ($v_a$) are not relativistic, the three velocity fields are related by the Lorentz transformation~\cite{M}
\begin{equation}
\tilde{u}_a= u_a+ v_a\,,  \label{Lorentz}
\end{equation}
Given that both frames are defined at the same spacetime event, the above has also been used to relate the various kinematic and matter variables evaluated in the two coordinate systems.

In studies of linear peculiar velocities, the key relations are between the energy fluxes ($q_a$) and the 4-accelerations ($A_a$). For zero pressure, we have
\begin{equation}
\tilde{q}_a= q_a- \rho v_a \hspace{10mm} {\rm and} \hspace{10mm} \tilde{A}_a= A_a+ \dot{v}_a+ Hv_a\,,  \label{lrels1}
\end{equation}
on an Einstein-de Sitter background~\cite{M}. Here, $\rho$ and $H$ are the background density and Hubble parameter respectively. Also, the tilded variables are in the matter frame and the non-tilded in that of the CMB. Following~\cite{T1}, as well as~\cite{M}, we set the flux and the 4-acceleration to zero in the matter frame. This implies nonzero flux ($q_a=\rho v_a$) and nonzero 4-acceleration ($A_a=-\dot{v}_a-Hv_a$) in the coordinate system of the CMB. The latter relation recasts as
\begin{equation}
\dot{v}_a+ Hv_a= -A_a  \label{vdot}
\end{equation}
and provides the linear evolution formula of the peculiar-velocity field. Hence, the 4-acceleration is the sole driving force of linear peculiar-velocity perturbations, which makes $A_a$ the key agent that determines their evolution. Equation (\ref{vdot}) is the starting point in both the quasi-Newtonian and the relativistic treatments. However, the agreement stops here and the two studies subsequently diverge because they employ different forms of 4-acceleration.\vspace{-10pt}

\subsection{The quasi-Newtonian approach}\label{ssQNA}\vspace{-5pt}
Setting the linear vorticity to zero, allows one to express the 4-acceleration as the gradient of a scalar. Then, the quasi-Newtonian 4-acceleration is monitored by the set~\cite{M}
\begin{equation}
A_a= {\rm D}_a\varphi \hspace{10mm} {\rm and} \hspace{10mm} \dot{\varphi}= -\Theta/3\,,  \label{QNAa}
\end{equation}
where $\varphi$ is the scalar. However, $\varphi$ is arbitrary, with no explicit expression specifying it and is indistinguishable from the Newtonian gravitational potential for all practical purposes. In addition, the evolution equation (\ref{QNAa}b), which also requires the linear shear to vanish, is a non-uniquely specified ansatz. Taking the time derivative of (\ref{QNAa}a) and combining it with (\ref{QNAa}b), one arrives at the following propagation formula
\begin{equation}
\dot{A}_a+ 2HA_a+ {3\over2}H^2v_a= 0\,,   \label{QNdotA}
\end{equation}
for the linear evolution of the quasi-Newtonian 4-acceleration.\footnote{The system (\ref{QNAa}) was the starting point in the quasi-Newtonian study of~\cite{M}. In~\cite{CM}, the authors begun their calculation from (\ref{QNdotA}). However, the latter follows directly from the former and it is also non-uniquely specified.} Combining (\ref{vdot}) with the above leads to the differential equation
\begin{equation}
\ddot{v}_a+ 3H\dot{v}_a- H^2v_a= 0\,, \label{ddotv}
\end{equation}
which on an Einstein-de Sitter background (where $H=2/3t$) accepts the solution
\begin{equation}
v_a= \mathcal{C}_1t^{1/3}+ \mathcal{C}_2t^{-4/3}\,.  \label{Nlv}
\end{equation}
The above result is identical to that of the purely Newtonian analysis. This was inevitable since the formulae leading to solution (\ref{Nlv}) are identical to their Newtonian counterparts.

Overall, the quasi-Newtonian approximation monitors the linear evolution of peculiar velocities solely through Eqs.~(\ref{vdot}) and (\ref{QNdotA}). However, this system has been derived only after imposing the Newtonian-like expression (\ref{QNAa}a) for the 4-acceleration and after adopting the ansatz (\ref{QNAa}b) for its evolution. Consequently, the quasi-Newtonian and the purely Newtonian approaches rely on identical systems of differential equations, which inevitably lead to identical solutions. As expected, both treatments bypass the gravitational contribution of the peculiar flux.

By contrast, in the proper relativistic approach, the 4-acceleration arises naturally from the energy-density conservation law, without the need to impose any linear constraints. Moreover, the relativistic analysis involves a third propagation formula, derived from the Raychaudhuri equation. As a result, this framework consistently incorporates the gravitational effect of the peculiar flux, leading to qualitatively different solutions (see \S~\ref{ssRA} below).\vspace{-10pt}

\subsection{The relativistic approach}\label{ssRA}\vspace{-5pt}
Just like the quasi-Newtonian analysis, the relativistic treatment also sets the peculiar flux to zero in the matter frame, which therefore follows timelike geodesics (i.e.~$\tilde{q}_a=0= \tilde{A}_a$). Also, both studies monitor the peculiar-velocity evolution in the CMB frame. However, the relativistic approach imposes none of the quasi-Newtonian constraints. The essential physical difference between the two studies lies in the fact that the proper relativistic treatment consistently incorporates the gravitational contribution of the peculiar flux. This leads to additional flux-terms in the evolution equations of the linear peculiar-velocity field. Among them, is the energy conservation law ($\dot{\rho}=-\Theta\rho-{\rm D}^aq_a$), the linear gradient of which yields the expression
\begin{equation}
A_a= -{1\over3H}\,{\rm D}_a\vartheta- {1\over3aH}\left(\dot{\Delta}_a+\mathcal{Z}_a\right)\,,  \label{RAa}
\end{equation}
for the 4-acceleration~\cite{TT}. Here, the gradients $\Delta_a$ and $\mathcal{Z}_a$ describe inhomogeneities in the matter density and the expansion respectively, while $\vartheta = {\rm D}^a v_a$ is the 3-divergence of the peculiar-velocity field. Hence, the right-hand side contains both the contribution of the peculiar flux and the effects of structure formation. Equation (\ref{RAa}) provides the proper relativistic expression for the linear 4-acceleration, in contrast to the quasi-Newtonian relation (\ref{QNAa}a). Clearly, had the energy-conservation law been employed in~\cite{M}, that study could also have arrived at the relativistic expression (\ref{RAa}) for the 4-acceleration, instead of the Newtonian-like (\ref{QNAa}a).

The form of the relativistic 4-acceleration implies the need for an additional evolution equation, namely the propagation of the expansion gradient $\mathcal{Z}_a$. This follows by taking the gradient of the Raychaudhuri equation, which at the linear level reads
\begin{equation}
\dot{\mathcal{Z}}_a= -2H\mathcal{Z}_a- {1\over2}\,\rho\Delta_a- {9\over2}\,aH^2A_a+ a{\rm D}_a{\rm D}^bA_b\,.  \label{ldotcZ}
\end{equation}

The profound differences between the quasi-Newtonian and the fully relativistic formulations naturally lead to corresponding differences in their results. Indeed, combining Eqs.~(\ref{vdot}), (\ref{RAa}) and (\ref{ldotcZ}) yields the differential equation~\cite{T1}
\begin{equation}
\ddot{v}_a+ H\dot{v}_a- {3\over2}\,H^2v_a= {1\over3aH}\left(\ddot{\Delta}_a+ 2H\dot{\Delta}_a- {3\over2}\,H^2\Delta_a\right)\,. \label{lRddotv}
\end{equation}
This is the correct relativistic equation monitoring the linear evolution of peculiar velocities on an Einstein-de Sitter background, in contrast to the quasi-Newtonian expression (\ref{ddotv}).

The left-hand side of (\ref{lRddotv}) consists of terms involving the peculiar velocity and its derivatives, whereas the right-hand side involves the density gradient and its derivatives, each with both temporal and spatial dependence. Consequently, Eq.~(\ref{lRddotv}) is a non-homogeneous differential equation and subject to the related standard theorems. Among these is a well-known theorem that provides mathematically rigorous and physically insightful solutions without having to solve the full equation. The theorem, which was not mentioned in~\cite{CM}, but is readily available in all related undergraduate textbooks, states:

\textit{The general solution of a non-homogeneous differential equation is obtained by adding to the general solution of its homogeneous component any particular solution of the full equation.}

Applying the above mathematical theorem, we may isolate the homogeneous left-hand side of Eq.~(\ref{lRddotv}), which admits the solution~\cite{T1}
\begin{equation}
v= \mathcal{C}_1t+ \mathcal{C}_2t^{-2/3}\,,  \label{Rlv}
\end{equation}
with the dominant mode ($v \propto t$) growing faster than its Newtonian/quasi-Newtonian counterpart. This mode also sets the minimum linear growth-rate of the peculiar-velocity field within the relativistic framework. Any particular solution of the full non-homogeneous equation (\ref{lRddotv}) would alter the physical behaviour only if it grows faster than the $v\propto t$ mode.

There is a clear physical interpretation underlying the mathematics that led to solution (\ref{Rlv}). Isolating and solving the homogeneous part of Eq.~(\ref{lRddotv}) amounts to neglecting the effects of the density and expansion gradients, while retaining those of the peculiar flux. Setting $\Delta_a$ and $\mathcal{Z}_a$ to zero in Eqs.~(\ref{RAa}), (\ref{ldotcZ}) and then combining the resulting expressions, one finds that $A_a=$~constant~\cite{PT}. Substituting a constant 4-acceleration into Eq.~(\ref{vdot}) reproduces solution (\ref{Rlv}).

Accordingly, in the absence of inhomogeneity effects, the scaling $v \propto t$ corresponds to the linear growth-rate of peculiar-velocity perturbations on an Einstein–de Sitter background. When such effects are included, however, the $v \propto t$ mode represents the minimum linear growth-rate. This highlights the physical significance of the mathematical theorem stated above.

Before reviewing the criticisms of the relativistic treatment, as presented in~\cite{CM}, we first provide a critique of the quasi-Newtonian approach. From a mathematical standpoint, the latter relies on a simpler system of two differential equations (see Eqs.~(\ref{vdot}) and (\ref{QNdotA})). This simplicity, however, is not intrinsic. Rather, it was enforced by adopting the Newtonian-like expression (\ref{QNAa}a) for the 4-acceleration and by introducing the ansatz (\ref{QNAa}b). These, in turn, follow from the strict quasi-Newtonian constraints imposed upon the linear vorticity and shear.

By contrast, the relativistic analysis employs a fundamentally different system of three equations (see Eqs.~(\ref{vdot}), (\ref{RAa}) and (\ref{ldotcZ})), all of which arise naturally from Einstein's equations. Physically, this distinction reflects the fact that only the relativistic study incorporates the gravitational input of the peculiar flux. The limitations of the quasi-Newtonian approach are therefore both mathematical and physical, though neither aspect has been recognised in~\cite{CM}.

The underlying physical differences between the two approaches are clearly manifested in their respective forms of the 4-acceleration (compare (\ref{QNAa}a) with (\ref{RAa})). In the quasi-Newtonian treatment, peculiar velocities are driven by the gradient of an arbitrary (Newtonian-like) scalar. In contrast, the relativistic driving force results from structure formation and the associated inhomogeneities. The reader may judge themselves which of the two approaches has better credentials.\vspace{-10pt}

\subsection{Reviewing the criticisms}\label{ssRCs1}\vspace{-5pt}
Let us begin by noting again that, had the gradient of the energy-conservation law been employed in~\cite{M}, that study could also have arrived at expression (\ref{RAa}) for the 4-acceleration. Then, this entire exchange would have been completely unnecessary.\vspace{-6.25pt}

\begin{itemize}
\item The relativistic analysis of~\cite{T1}, adopts the setup of the quasi-Newtonian treatment of~\cite{M}, in order to facilitate the direct comparison of the two studies. Both approaches set the flux to zero in the matter frame, which follows timelike geodesics (i.e.~$\tilde{q}_a=0= \tilde{A}_a$). Also, both use the CMB frame to monitor the evolution of the peculiar-velocity field. The two studies diverge solely because they employ different forms of 4-acceleration. The quasi-Newtonian 4-acceleration (\ref{QNAa}a) is essentially an ansatz, whereas the relativistic (\ref{RAa}) follows naturally from the energy conservation law. The latter was never involved in the quasi-Newtonian calculations and this omission reduced the analysis to purely Newtonian.\vspace{-6.25pt}

\item It is claimed in~\cite{CM} that the quasi-Newtonian system of Eqs.~(\ref{vdot}) and (\ref{QNdotA}) is a closed subset of the relativistic system (\ref{vdot}), (\ref{RAa}) and (\ref{ldotcZ}). This is not so. If it were, one would have been able to obtain the former set from the latter, as well as Eq.~(\ref{ddotv}) from (\ref{lRddotv}), without imposing the quasi-Newtonian constraints. The authors are invited to prove otherwise.\vspace{-6.25pt}

\item The authors admitted that Eq.~(\ref{lRddotv}) is correct, but claimed a discrepancy between (\ref{lRddotv}) and the differential formula
    \begin{equation}
    \ddot{v}_a+ {1\over2}\,Hv_a- 2H^2v_a= {1\over3H}\,{\rm D}_a\dot{\vartheta}+ {1\over3aH}\left(\ddot{\Delta}_a+\dot{\mathcal{Z}}_a\right)\,, \label{Rddotv1}
    \end{equation}
    derived in the original relativistic study of~\cite{TT}. However, the two equations are not in conflict, nor is one a mere re-arrangement of the other, as implied in~\cite{CM}. Instead, Eq.~(\ref{lRddotv}) is the improvement/refinement of (\ref{Rddotv1}). Indeed, one can readily obtain (\ref{lRddotv}) by simply substituting $\dot{\mathcal{Z}}_a$ from Eq.~(\ref{ldotcZ}) into (\ref{Rddotv1}). This should not have been so difficult to recognise.\footnote{For readers interested in the underlying physics connecting the differential formulae (\ref{lRddotv}) and (\ref{Rddotv1}), see~\cite{PT}.}\vspace{-6.25pt}

\item In~\cite{CM}, the authors claim that one cannot isolate the homogeneous left-hand side of (\ref{lRddotv}) because the peculiar velocity and the density gradients interact. This is a novel assertion that could revolutionize differential-equation theory and mathematical physics in general. However, unless the authors can rigorously (mathematically) prove their claim, rather than merely argue for it, (\ref{lRddotv}) is a non-homogeneous differential equation, and the above standard theorem applies.\vspace{-6.25pt}

\item To support their quasi-Newtonian case, the authors presented a metric-based treatment of linear peculiar velocities in Appendix~A. This was done using the Newtonian gauge. The gauge-choice, however, undermines rather than supports the authors' case. A mere look at the calculation in Appendix~A reveals that it simply reproduces the purely Newtonian calculation. The apparent agreement was thus fixed ab initio. It goes without saying that the genuinely relativistic gravitational input of the peculiar flux is nowhere to find.

    Incidently, the authors have inadvertently illustrated that, at least in studies involving peculiar velocities, the Newtonian gauge is as ill-suited as the quasi-Newtonian approach.\vspace{-6.25pt}

\item An additional criticism was that the numerical simulations of structure formation appear to agree with the Newtonian picture. However, numerical results are tested against the analytical. In addition, to the best of our knowledge, there are no genuinely relativistic simulations specifically addressing cosmological peculiar velocities and their evolution. Moreover, the available simulations are typically carried out in a specific gauge - often in the Newtonian gauge - where any potential analysis of peculiar velocities de facto reduces to purely Newtonian (see above).\vspace{-6.25pt}

\item The authors claim that the relativistic solution produces large residual peculiar velocities. However, the magnitude of the residuals depends primarily on the chosen initial conditions. 
    
    At the same time, unlike the Newtonian and quasi-Newtonian approaches, the relativistic analysis offers theoretical support for a growing number of surveys that report bulk flows much faster and much deeper than previously anticipated. The Newtonian and quasi-Newtonian frameworks are unable to account for such findings. Among these surveys are those indicating a late-time suppression and decay of the peculiar-velocity field~\cite{CMSS}. The relativistic results can readily accommodate this late-time behaviour~\cite{T1,TPA}, in clear contrast to the Newtonian and quasi-Newtonian predictions, even when starting from comparatively weaker initial conditions.\vspace{-6.25pt}

\item Finally, in their manuscript the authors refer to the growing velocity mode as spurious, bypassing the fact that the relativistic analysis is fully gauge-invariant and there are no spurious modes in the solution. Spurious modes are more likely to appear in gauge-depended studies, like the one given in the Appendix of~\cite{CM}.
\end{itemize}\vspace{-6.25pt}

In~\cite{CM}, the authors make no reference to the concerns and cautionary remarks already expressed in~\cite{EMM}, and defend their quasi-Newtonian work solely on the basis of the consistency of the imposed constraints. Be that as it may, the central point remains: the quasi-Newtonian analysis - at least as applied to peculiar velocities (and perhaps more generally) - is not relativistic and should not be presented as such.\vspace{-5pt}

\section{Bulk flows and the deceleration parameter}\label{sBFDP}\vspace{-5pt}
Looking back to the history of astronomy, one can find several examples where relative motions have interfered and contaminated the observations, which in a number of cases led to gross and long-lasting misinterpretations. Cosmology is not a priori immune to such contaminations.\vspace{-10pt}

\subsection{Unsuspecting vs informed bulk-flow observers}\label{ssBFCMBF}
\begin{figure}[!tbp]
  \begin{subfigure}[b]{0.475\textwidth}
    \includegraphics[width=\textwidth]{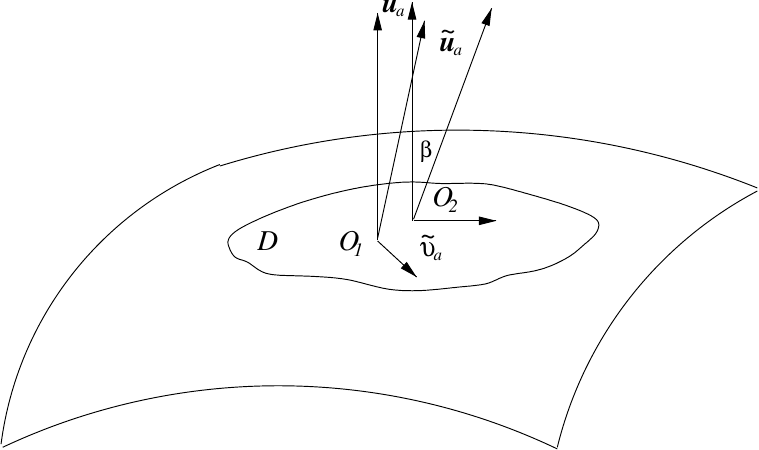}
    \caption{Bulk-flow \& CMB frames.}
    \label{fig:f1}
  \end{subfigure}
  \hfill
  \begin{subfigure}[b]{0.475\textwidth}
    \includegraphics[width=\textwidth]{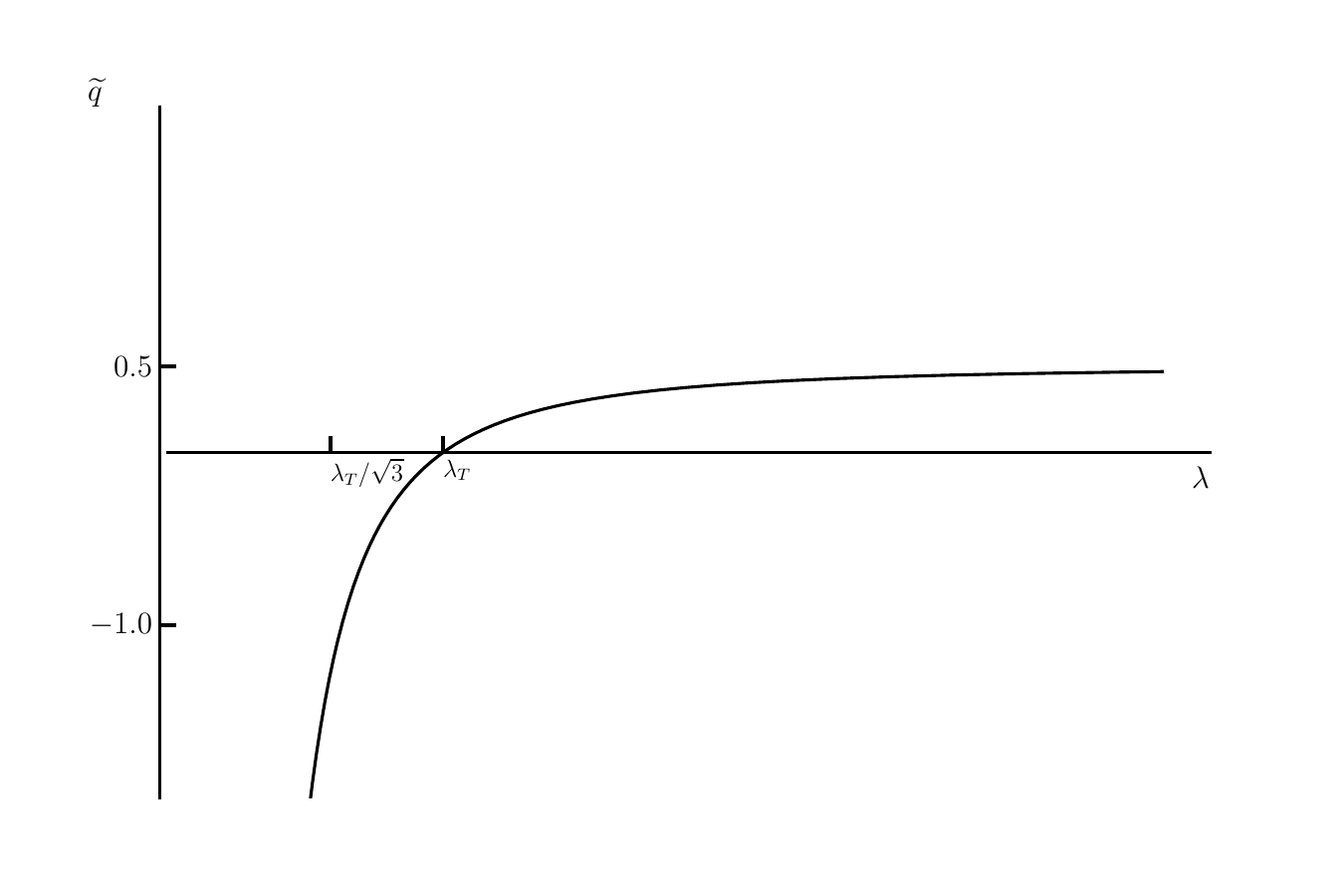}
    \caption{The $\tilde{q}$-profile.}
    \label{fig:f2}
  \end{subfigure}\vspace{-5mm}
\caption{(a) Observers ($O_1$, $O_2$) inside a bulk-flow ($D$), moving with 4-velocity $\tilde{u}_a$ and having mean bulk velocity $v_a$ relative to their CMB counterparts (with 4-velocity $u_a$). (b) The profile of the deceleration parameter measured in the bulk-flow frame, ``contaminated'' by the bulk-flows local contraction (see Eq.~(\ref{tq3})). The Einstein-de Sitter limit ($\tilde{q}\rightarrow q=0.5$) is recovered at sufficiently high redshifts, where $\lambda\gg\lambda_T$. Inside the transition length ($\lambda_T$), $\tilde{q}$ appears to turn negative and even to cross the phantom divide ($\tilde{q}=-1$) at $\lambda=\lambda_T/\sqrt{3}$.}  \label{fig:qplot}
\end{figure}

\vspace{-5pt}
Consider a large scale bulk flow moving relative to the CMB. At every point inside the bulk-flow there are two coordinate systems at all times (see Fig.~\ref{fig:f1}). The $\tilde{u}_a$-frame of a typical bulk-flow observer and that of the CMB ($u_a$), related by the Lorentz transformation (\ref{Lorentz}).

Observers in relative motion generally disagree in their measurements, but one can relate the various kinematical, geometrical and matter variables by means of (\ref{Lorentz}). Some of these relations were given in Eqs.(2.8)-(2.11) of~\cite{CM}. Here, we will only need the first, namely
\begin{equation}
\tilde{\Theta}= \Theta+ {\rm D}^av_a+ A^av_a  \label{Thetas1}
\end{equation}
which relates the cosmic expansion rates ($\tilde{\Theta}$ and $\Theta$) in the two frames. On a Friedmann background, the last term in the above is nonlinear, so the linearised version of (\ref{Thetas1}) reads
\begin{equation}
\tilde{\Theta}= \Theta+ \vartheta\,, \hspace{10mm} {\rm with} \hspace{10mm} \tilde{\Theta}^{\prime}= \dot{\Theta}+ \dot{\vartheta} \label{Thetas2}
\end{equation}
where $\vartheta={\rm D}^av_a$. Note that primes indicate time derivatives in the bulk-flow frame of the matter and overdots in that of the CMB. According to (\ref{Thetas2}), the expansion rates and their time derivatives differ and their difference is entirely due to relative-motion effects. Then, in locally contracting bulk flows (with $\vartheta<0$), the expansion rate is smaller than that of the surrounding universe. However, the effect is weak since $|\vartheta|/\Theta\ll1$ to linear order. The ratio $|\dot{\vartheta}|/\dot{\Theta}$, on the other hand, is not necessarily small.

Since $\tilde{\Theta}\neq\Theta$ and $\tilde{\Theta}^{\prime}\neq\dot{\Theta}$, the deceleration parameters in the two frames should differ as well. Indeed, by definition, we have
\begin{equation}
\tilde{q}= -\left(1+{3\tilde{\Theta}^{\prime}\over\tilde{\Theta}^2}\right) \hspace{10mm} {\rm and} \hspace{10mm} q= -\left(1+{3\dot{\Theta}\over\Theta^2}\right)\,,  \label{qs}
\end{equation}
in the bulk-flow and the CMB frames respectively. Combining (\ref{Thetas2}) and (\ref{qs}) and keeping up to linear-order terms, leads to the following relation
\begin{equation}
\tilde{q}= q- {\dot{\vartheta}\over3H^2}\,,  \label{tq1}
\end{equation}
between the two deceleration parameters. Therefore, $\tilde{q}\neq q$ and their difference is entirely due to relative-motion effects (as expected). On using (\ref{vdot}) and (\ref{RAa}), Eq.~(\ref{tq1}) recasts into~\cite{T2}
\begin{equation}
\tilde{q}= q+ {1\over9}\left({\lambda_H\over\lambda}\right)^2 {\vartheta\over H}\,,  \label{tq2}
\end{equation}
where $\lambda_H$ is the Hubble scale and $\lambda$ is the characteristic bulk-flow scale. Hence, the relative-motion effect on the local deceleration parameter ($\tilde{q}$) is scale-dependent and fades away with increasing scale (as expected). Well inside the Hubble horizon, on the other hand, the relative-motion effect can dominate despite the fact that $|\vartheta|/H\ll1$ always. Crucially, the overall effect also depends on whether the bulk flow is locally expanding or contracting. In the latter case, $\vartheta<0$ and the sign of the local deceleration parameter ($\tilde{q}$) can turn negative, while $q$ remains positive.

One can calculate the ``transition scale'' ($\lambda_T$), where $\tilde{q}$ appears to change from positive to negative, by setting the right-hand side of (\ref{tq2}) to zero. Then, $\lambda_T= \sqrt{|\vartheta|/9q}\lambda_H$ and Eq.~(\ref{tq2}) reads~\cite{T2}
\begin{equation}
\tilde{q}= q\left[1- \left({\lambda_T\over\lambda}\right)^2\right]\,,  \label{tq3}
\end{equation}
in locally contracting bulk flows. The second term on the right-hand side of the above, as well as those in (\ref{tq1}) and (\ref{tq2}), quantify the ``contamination'' of $\tilde{q}$ by the bulk flow's local contraction. As expected, the contamination is scale-depended, fading away on large scales (so that $\tilde{q}\rightarrow q$ when $\lambda\gg\lambda_T$) and dominating on scales smaller than $\lambda_T$ (where $\tilde{q}<0$ -- see Fig.~\ref{fig:f2}).

Let us turn to the physical implications of Eq.~(\ref{tq3}) and of Fig.~\ref{fig:f2}. All observers comoving with the bulk flow share a common rest-frame, have access to the same observational data and therefore perform identical measurements. Nonetheless, there are two distinct groups of bulk-flow observers. The first are the \textit{unsuspecting observers}, who are either unaware of their location within a locally contracting bulk flow, or neglect to account for it. The second group are the \textit{informed observers}, who are fully aware of their location and incorporate this into their interpretation. Consequently, although all bulk-flow observers - unsuspecting and informed alike - measure the same value of $\tilde{q}$, they differ in the interpretation of their measurement.

Without accounting for their peculiar motion, the unsuspecting observers unwittingly identify themselves with their CMB partners and remain unaware of the contamination term in Eq.~(\ref{tq3}). They are therefore inclined to believe that $\tilde{q}=q$, namely that they have measured the true deceleration parameter of the universe. In contrast, the informed observers recognize that $\tilde{q}\neq q$, and that the local deceleration parameter has been contaminated by their relative motion and by the fact that they reside within a contracting bulk flow.

When the unsuspecting observers examine Fig.~\ref{fig:f2}, they “see” a universe that was decelerating in the past but began accelerating relatively recently. Their informed counterparts, however, understand that this apparent behavior may simply be an artefact caused by the local contraction of their bulk flow. In other words, the unsuspecting observers have misinterpreted a local effect as evidence of a recent global transition. More specifically, they have mistaken the local contraction of their own bulk flow for global acceleration of the surrounding universe.

\textit{Intuitively, this situation is akin to passengers in a car that is slowing down, who perceive the other vehicles as accelerating away. The idea is conceptually very simple, though recognizing it requires a certain degree of physical insight.}

Is there a way for the bulk-flow observers to test whether their measurements have been biased by their own motion? The answer is affirmative. If the observed cosmic acceleration is indeed an artefact of the observers’ peculiar flow, the data should exhibit the characteristic “trademark” signature of relative motion. Specifically, this would manifest as an apparent (Doppler-like) dipolar anisotropy in the sky-distribution of the deceleration parameter. \textit{To the bulk-flow observers, the universe would appear to accelerate more rapidly along a particular direction in the sky and, correspondingly, more slowly in the opposite (antipodal) direction. Moreover, the amplitude of this dipole should decrease with increasing redshift~\cite{T2}.}

Over the past fifteen years or so, many studies have claimed the possible presence of such a dipolar anisotropy in the supernova data. It was not until~\cite{CMRS}, however, that the dipole identified in the JLA catalogue was attributed to our location within a contracting bulk flow. A few years later, a similar dipole was detected in the Pantheon+ dataset, notably with a magnitude that decayed with redshift~\cite{SRST}. Around the same period, a complementary analysis by~\cite{CRMC} reported a dipolar anisotropy in the distribution of $\Omega_{\Lambda}$, also consistent with being an artefact of our peculiar motion (as suggested/predicted in~\cite{T2}). It is still early days, however, since we need more and better data. Hopefully, these will become available in the near rather than the distant future.\vspace{-10pt}

\subsubsection{Reviewing the criticisms}\label{sssRCs2}\vspace{-5pt}
\begin{itemize}
\item In comparing the matter and CMB frames, the authors elevate the former while relegating the latter to an “auxiliary” role. This is difficult to justify. The universe contains a rich hierarchy of large-scale structures, such as superclusters, voids, and bulk flows that may be locally expanding or contracting. Each of these structures defines its own local matter frame. Consequently, cosmological measurements performed within different matter frames are inevitably ``contaminated'' by both the observer’s location and the local kinematics.

    By contrast, the CMB frame is commonly identified with the rest-frame of the universe, under the standard assumptions that the observed dipole is kinematical in origin and the cosmological principle holds. Although different observers may measure different dipoles in their local CMB sky, they can all infer and agree upon the same underlying CMB frame. In this sense, the CMB frame is not merely auxiliary, but both unique and global.\vspace{-5pt}

\item There are also direct inconsistencies in~\cite{CM}. For instance, the authors criticize relations (\ref{tq1})-(\ref{tq3}) on the grounds that variables cannot be related across different congruences. This is a striking claim, given that for over a century such relations have been routinely established through Lorentz transformations. Indeed, the authors themselves employed this very procedure in deriving the linear relation (\ref{Thetas1}), along with a long list of others~\cite{M}.

    Thus, on the one hand, they accept relation (\ref{Thetas1}), connecting the cosmic expansion rates, while on the other they reject relations (\ref{tq1})-(\ref{tq3}) linking the corresponding deceleration parameters, despite the fact that the latter follow by simply taking the time-derivative of the former. This constitutes a clear internal contradiction, or, to borrow the Greek expression, a ``$\sigma\chi\acute{\eta}\mu\alpha$ $o\xi\acute{\upsilon}\mu\omega\rho o\nu$''.\vspace{-5pt}

\item This is not the only internal contradiction. Toward the middle of \S~4 in~\cite{CM} - see Eqs.~(4.2)-(4.4) there - the authors presented a brief calculation intended to demonstrate that $\tilde{q}$ is always positive. However, that calculation contains no peculiar velocities whatsoever - they are entirely absent. This happens because the calculation takes place exclusively in the bulk-flow frame, where the peculiar-velocity effects have already been switched off (recall that $\tilde{q}_a=0=\tilde{A}_a$ there), without involving at all the CMB frame, where $q_a,\,A_a\neq0$ and the relative-motion effects are accounted for. In effect, the authors first removed the peculiar velocities from their calculation and then ``proved'' that they have no impact on $\tilde{q}$. This line of reasoning is difficult to take seriously.\footnote{The authors could have accounted for the peculiar velocities by also involving the CMB frame in their calculation. Their failure to do so likely explains why the CMB frame was relegated to a merely auxiliary role in the first place.}

    Incidently, relations (\ref{tq1})-(\ref{tq3}) do not imply that the local deceleration parameter has genuinely changed sign. Rather that it appears to do so, and from the perspective of the unsuspecting observers only. While such misinterpretations are not uncommon, even in everyday contexts, recognizing them requires a certain level of physical insight.

    That said, there is an unintended positive aspect. In their confused reasoning, the authors have unwittingly identified themselves with the unsuspecting observers and inadvertently illustrated how they think and operate. Despite living in a universe that exhibits large-scale peculiar motions - motions that they do not incorporate into their analysis - the authors remain confident in the completeness and robustness of their conclusions.\vspace{-10pt}
\end{itemize}

\section{Conclusions and discussion}\label{sCD}\vspace{-5pt}
The relativistic studies of linear peculiar velocities by~\cite{TT,T1} were not the first to criticize the quasi-Newtonian approach. As mentioned above, its limitations had already been noted and criticised in~\cite{EMM}. The relativistic analyses confirmed the concerns raised in~\cite{EMM}, clarified the extent of the problem, and demonstrated how it can be resolved.

It is natural and understandable that the authors seek to defend their earlier work on the quasi-Newtonian approximation~\cite{M}. However, there is also a matter of common sense to consider. We are faced with two distinct approaches. The first - the quasi-Newtonian - sets both vorticity and shear to zero, eliminates gravitational waves, adopts Newtonian-like expressions for the 4-acceleration, and neglects the contribution of the peculiar flux to the gravitational field. The second is a fully relativistic treatment, which imposes no linear constraints, derives the 4-acceleration directly from the relativistic conservation laws, and consistently incorporates the gravitational effects of the peculiar flux. To claim that the former is mathematically more rigorous, physically more transparent and overall superior to the latter, stretches credibility.

\textit{Ultimately, however, had the gradient of the energy-conservation law been employed in~\cite{M}, that study could also have arrived at the proper relativistic expression (\ref{RAa}) for the 4-acceleration. In that case, the ensuing exchange between the quasi-Newtonian and relativistic approaches would have been entirely superfluous and therefore completely unnecessary (see also \S~\ref{ssRA}).}

The CMB frame has long been identified with the rest-frame of the universe. This identification relies on the implicit assumptions that the CMB dipole is of kinematical origin and that the cosmological principle holds. Under these conditions, the CMB provides the natural cosmic reference frame for studying large-scale peculiar motions and their implications~\cite{EMM}. Consequently, measurements performed locally in various matter frames do not necessarily coincide with those made in the cosmic reference system, primarily due to relative-motion contaminations. These include the measurements of the cosmic expansion-rate, as well as those of the deceleration/acceleration of the universe. In both cases, one can identify contaminations arising from relative motion, which can lead unsuspecting observers to misinterpret the data and draw erroneous conclusions.

Admittedly, the suggestion that the late-time acceleration of the universe might be nothing more than a deception caused by our own peculiar motion is difficult for most cosmologists to accept. If true, it would imply that for more than a quarter of a century we have been unsuspecting observers who unwittingly misinterpreted the local contraction of their bulk flow as evidence for the acceleration of the surrounding universe. A negative reception is therefore to be expected, especially since some “unsuspecting observers” are likely to act in ``confederacy''~\cite{S}. Yet this would not be the first instance in which relative motions have led to profound misinterpretations of reality - errors that have at times persisted for centuries, if not millennia. History reminds us that nature often appears far simpler when viewed from the proper perspective.

\end{document}